# All-MOCVD-Grown Gallium Nitride Diodes with Ultra-Low Resistance Tunnel Junctions

Syed M. N. Hasan[1], Brendan P. Gunning[2], Zane J.-Eddine[1], Hareesh Chandrasekar[1], Mary H. Crawford[2], Andrew Armstrong[2], Siddharth Rajan[1], and Shamsul Arafin[1]*

[1] Department of Electrical and Computer Engineering, The Ohio State University, Columbus, OH 43210, U.S.A.

[2] Sandia National Laboratories, Albuquerque, NM 87185, U.S.A.

E-mail: arafin.1@osu.edu

**Abstract**

We carefully investigate three important effects including postgrowth activation annealing, delta (δ) dose and p+-GaN layer thickness and experimentally demonstrate their influence on the electrical properties of GaN p-n homojunction diodes with a tunnel junction (TJ)-based p-contact. The p-n diodes and TJ structures were monolithically grown by metalorganic chemical vapor deposition (MOCVD) in a single growth step. By optimizing the annealing time and temperature for magnesium (Mg) activation and introducing δ-doses for both donors and acceptors at TJ interfaces, a significant improvement in electrical properties is achieved. For the continuously-grown, all-MOCVD GaN homojunction TJs, ultra-low forward voltage penalties of 158 mV and 490 mV are obtained at current densities of 20 A/cm$^2$ and 100 A/cm$^2$, respectively. The p-n diode with an engineered TJ shows a record-low normalized differential resistance of $1.6 \times 10^{-4}$ Ω-cm$^2$ at 5 kA/cm$^2$.

# 1. Introduction

Since the successful demonstration of p-GaN as an active layer [1], GaN-based devices have drawn significant attention in the area of optoelectronics [2, 3] and power electronics [4, 5]. While III-nitride light emitting diodes (LEDs) have been commercially available, a long-standing challenge of reduced external quantum efficiency at higher current levels remains, which limits the usability of LEDs for high intensity applications. One way to circumvent the efficiency droop issue is to use multiple active region LEDs connected by tunnel junctions (TJs) [6]. To achieve this, it is critical to develop TJs that have low added voltage penalty across them to reduce electrical losses in such cascaded LEDs. While there have been significant efforts made over the last decade in developing low-resistance TJs [7], there are still important challenges associated with achieving low voltage-drop TJs using metal organic chemical vapor deposition (MOCVD). In this experimental study, we report the reduction of the voltage penalty in all-MOCVD GaN TJs by tailoring the p-doped region.

One of the major challenges for achieving low-resistivity GaN TJs via MOCVD is the activation of Mg dopants in the buried p-GaN layers. Hydrogen, which forms a stable complex with Mg acceptors, has low diffusivity through the n-type GaN layers [8, 9], and therefore, activation of p-type GaN through overlying n-type GaN layers is challenging. The well-known technique of high-temperature annealing at 725°C enables dissociation of the Mg-H complex, and diffusion of the H species laterally, enabling an activated p-GaN layer [10, 11]. Despite the reasonably good electrical performance seen in devices with small dimensions (<50 µm) using such a high-temperature process causes large-area devices to suffer from insufficient activation, resulting in higher voltage penalties [12-14]. The MOCVD growth technology-induced Mg-memory[15] and Mg built-up delay effects [16] introduce added challenges towards achieving the sharp doping profiles necessary for efficient GaN TJs. Nonetheless, an all-MOCVD process is highly preferred for its efficacy in terms of providing better-quality InGaN multi quantum well (MQW)-based active region, enabling high manufacturing throughput and cost-effective devices. Hence, further development of an activation annealing process for buried p-type Mg dopants and doping engineering within the diode structure is essential so that uniform and reproducible large hole concentrations can be obtained even in large-area devices.

Among several approaches, sidewall activation of Mg:GaN layers through an optimized postgrowth annealing process is one of the simple techniques to improve p- GaN conductivity. Introducing a delta ($\delta$) dose at the doped layer interfaces has also proven to be effective [17-19]. Compared to a continuously doped layer, the $\delta$-dose layer shows an order of magnitude higher carrier concentration. The reasons behind this improvement are reduced Mg self-compensation [18], reduction of surface defects [20] and Mg built up delay compensation [16]. Moreover, a polarization-engineered heterostructure by inserting a thin InGaN interlayer within the TJ shows great promise for interband tunneling [21, 22]. While the InGaN interlayer leads to excellent TJ performance, its presence in transparent GaN TJs may degrade the performance of cascaded LED structures through optical absorption. InGaN interlayer-free TJs is, therefore, an attractive approach.

Yuka et al. demonstrated H+ outdiffusion by exposing the sidewalls of the p-GaN layers after a mesa formation and annealing the sample at 725°C [12]. Recently, Akatsuka et al. demonstrated sidewall-activated GaN TJ with differential resistivity as low as $2.4 \times 10^{-4}$ Ω-cm$^2$ at 5 kA/cm$^2$ for a single run without any regrowth [23]. However, a controlled overlap between Mg and Si dopants reported in the work may affect the reproducibility of the device performance. MOCVD-grown TJs [3, 13, 14] exhibit higher voltage penalty compared to structures grown by PAMBE [24, 25], either as all-MBE or hybrid MBE/MOCVD structures [26-28], partly because MBE growth of p-GaN does not suffer from the Mg memory effect. Previous state-of-the-art results on MOCVD TJs [29-31] employed InGaN interlayers with relatively high indium (In) mole fractions to achieve low TJ voltage penalty ($<$ 0.1 V at 25 A/cm2). In fact, the major improvement made in terms of TJ efficiency is probably introducing a InGaN interlayer heterostructure, where the high polarization fields enable efficient tunneling across the junction [21, 22, 32].

In this work, we demonstrate homojunction, all-MOCVD TJs. We performed a detailed experimental study on the effects of postgrowth activation annealing and optimized this process towards improving the electrical properties of fully MOCVD-grown GaN p-n homojunction TJs. We also show that the sidewall activation process is critical to achieving low and repeatable voltage penalties across relatively large-area devices. Throughout the manuscript, two current density values, i.e. 20 A/cm2 and 100 A/cm2 are considered and the associated voltage penalty as a device performance metric is calculated. For blue LEDs, 20 A/cm2 is a widely accepted value since the peak EQE appears around that current density.

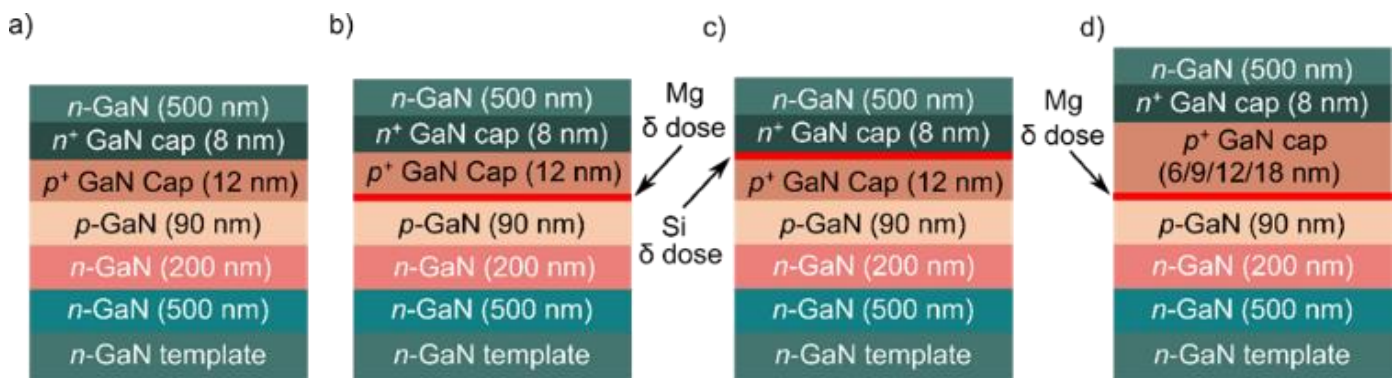

**Figure 1.** Schematic epitaxial layer structures of the samples with identical p-n diodes. Four samples including (a) TJ with no δ-dose (no-δ), (b) TJ with Mg δ-dose (Mg-δ), and (c) TJ with Si δ-dose (Si-δ), and (d) Mg-δ TJ with different p$^+$ GaN cap thickness were studied.

[28, 33] Voltage penalties of 158 mV at 20 A/cm2 and 490 mV at 100 A/cm2 are achieved, which represent some of the lowest values to-date for MOCVD-based single-step grown GaN p-n diodes.

## 2. Experimental Description

The p-n diode structure used in this study were grown by MOCVD in a Taiyo Nippon Sanso SR4000HT reactor at atmospheric pressure. C-plane sapphire with a micron-thick GaN template layer was used for the growth. The structure consists of 500-nm-thick n+-GaN for a bottom n-side ohmic contact, 200 nm n-GaN with a Si concentration of $2 \times 10^{16}$ cm$^{-3}$ and 90 nm p-GaN with a Mg concentration of $3 \times 10^{19}$ cm$^{-3}$. On top of the p-n junction, the TJs, comprised of two heavily p- and n-doped GaN layers with Mg and Si concentrations of $2 \times 10^{20}$ cm$^{-3}$ and $3 \times 10^{20}$ cm$^{-3}$, respectively, were grown. The growth temperature during the p+-GaN growth was 950°C. The growth was ended with a 500-nm-thick n-GaN layer to ensure good current spreading and to recover any growth defects. During the growth of the TJ, Si δ-dose was introduced between the p+ and n+ layer for sample "Si-δ", whereas for sample "Mg-δ", Mg δ-dose was introduced between the p and p+ layer, and one sample with a TJ without any δ-dose, referred to as "No- δ" was also grown. To investigate the Mg built-up delay effect, Mg-δ TJ samples with p+ GaN thickness 6 nm, 9 nm, 12 nm and 18 nm were processed together. Figure 1 schematically shows the layer structures of all the test samples used in this study.

After the materials growth, the samples were carefully examined under a standard optical microscope. Figure 2 (a) presents the Nomarski image of the MOCVD-grown *p-n* diode with TJ samples with no pits or defects. The AFM image, showing the surface morphology of the as-grown structures, is also presented in Fig. 2(c). The as-grown materials with a scan area of 10 μm × 10 μm exhibits a peak-to-peak variation of only ~7 nm and a root-mean-square (RMS) roughness value of 0.8 nm. For each sample, square mesas with 54 μm, 74 μm and 105 μm sides were defined by standard UV lithography and subsequent dry etching, down to the bottom $n^+$-GaN layer in a BCl$_3$/Cl$_2$/Ar gas mixture. Activation annealing was then performed using a rapid thermal annealing system in the temperature range of 900°-970°C. The annealing parameters, such as temperature and time, were first optimized using multiple Si-δ samples. A pure N$_2$ gas environment was used during annealing since this environment is proven to be sufficient for good electrical and optical performance of LEDs [34]. The *n*-side ohmic contact was then formed by depositing Ti/Al/Ni/Au (20/120/30/50 nm) followed by annealing at 850°C for 30 s. As for the top contact, Al/Ni/Au (30/30/150 nm) was deposited for TJ integrated *p-n* diodes (samples No-δ, Mg-δ and Si-δ). After the top contact deposition, the *p-n* diode was annealed at 500°C for 1 min in $N_2$ atmosphere.

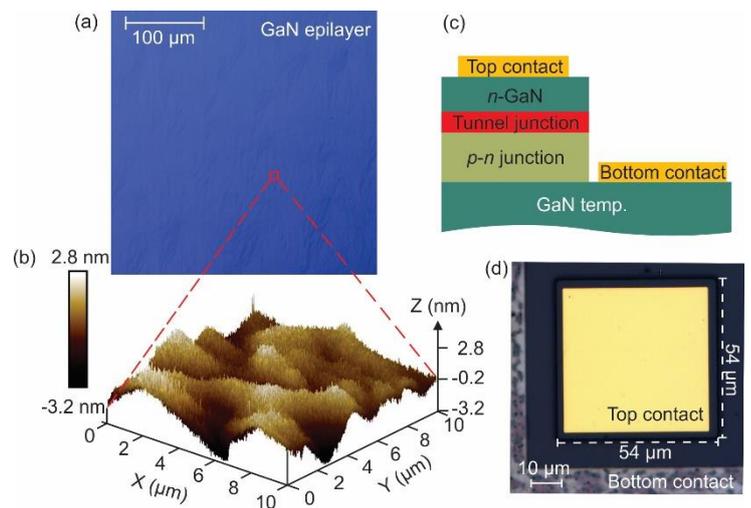

**Figure 2.** (a) Nomarski microscopy image, (b) Schematic cross-section, (c) 10 μm × 10 μm AFM image, and (d) optical microscope image of the fully-processed square-shaped test diodes.

## 3. Results and discussion

In order to improve the electrical properties of the all-MOCVD grown p-n diodes with a TJ, three major issues were addressed through this experimental study: post-growth annealing, delta (δ) dose and Mg built-up delay. In order to understand the three mechanism on the improvement of the diode electrical properties, it is important to determine the true voltage-drop across the TJs from the measured characteristics. For that, we first calculated the voltage drop for the p-n junction using the ideal diode equation (1). The series resistance in the equation includes sheet resistance at the doped epilayers and the top p-side n-contact resistance. The resistance due to the large-area bottom n-side contact on the highly-doped n-template layer was negligible.

$$J = q\left[\frac{D_n n_p}{L_n} + \frac{D_p p_n}{L_p}\right]\left(e^{q(V-Ir_s)/nkT} - 1\right) \quad (1)$$

$$\Delta V_{TJ} = V_1 - V \quad (2)$$

where Dn and Dp are the diffusion coefficients, Ln and Lp diffusion lengths, np and pn minority carrier density, V calculated voltage drop across diode without TJs using the diode equation and ideality factor η = 1 in Equation (1). V1 is the measured voltage drops for the p-n diode with TJ samples at the same current density. The TJ-only associated drop ΔVTJ was calculated from these two values represents the voltage penalty due to the TJ as itself. The voltage penalty for each device was measured at 20 A/cm$^2$ and 100 A/cm$^2$.

### 3.1 Side wall activation

For an effective thermal activation process, limited to lateral diffusion of H+ in p-GaN layer, a long annealing time and high temperature are needed [3, 12]. To prevent device degradation through interlayer diffusion, the activation temperature is usually kept below 750°C by employing a long annealing time [12, 14]. However, the long annealing time is still sometimes insufficient for large-area devices [3, 13]. Our experimental study shows that the electrical properties of MOCVD-grown p-n devices can be drastically improved by an optimized activation annealing process without sacrificing optical performance. In this experiment, we used the Si-δ samples (Fig. 1 (c)), which were activated under different conditions. Fig. 3(a) shows the plots of total voltage drop and differential resistance versus current density characteristics of the p-n junction prepared at various activation parameters including annealing temperature and time. As can be seen, there is a decrease in voltage drop due to reduced series resistance as the temperature increases from 900°C and/or the duration increases from 15 mins. The voltage drop for each device was measured at 20 and 100 A/cm$^2$ and plotted at Fig. 3(b). The voltage penalty variation across the devices is observed to be small and aligned with the voltage drop of samples annealed above 930°C, which shows 400 mV less voltage drop compared to the "900°C-15 min" sample at 100 A/cm$^2$. It is worth noting that no significant change in the reverse leakage current is observed, which indicates that the p-n diode remains unchanged in terms of electrical properties,

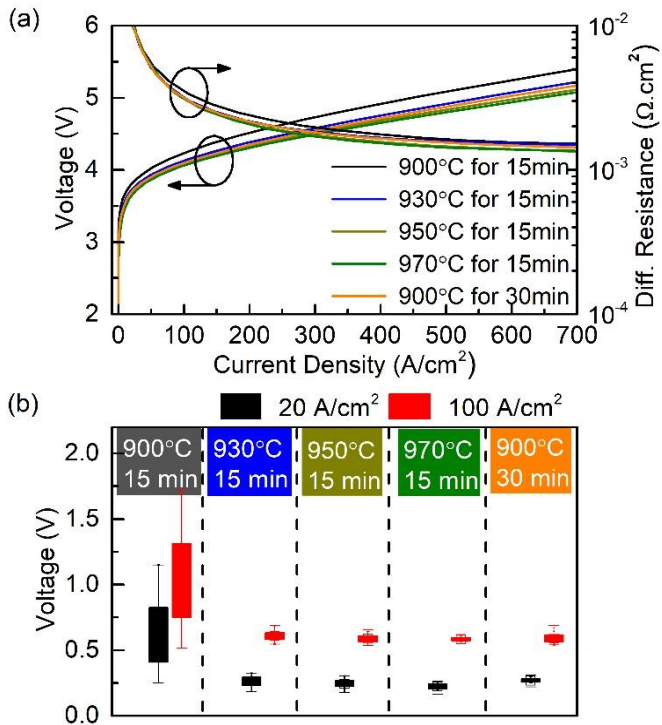

**Figure 3.** (a) Current density and differential resistance vs voltage characteristics and (b) comparison of voltage penalty of devices at 20 A/cm$^2$ and 100 A/cm$^2$ current densities activated in different annealing conditions.

resulting in minimum or no device degradation.

The diffusion of H+ has a square root relationship with the product of duration of the activation process and the diffusion coefficient, where the diffusion coefficient has an exponential dependence on temperature [14, 35]. As the n+-GaN layer acts as a blocking layer, preventing H+ from diffusing upward, the H+ ion diffusion in p-GaN layers is confined to the lateral direction. Once the p-GaN layer is completely activated, the electrical characteristics are expected to be independent of the device size. Fig. 4(a) shows the voltage penalty at 20 A/cm$^2$ as a function of the side-length of square-shaped devices. The voltage drop is nearly independent of the device size as the annealing time and temperature reach an optimum value, which is in good agreement with the physics discussed earlier. The overall voltage drop is found to be the lowest for devices annealed at 970°C-15 min. However, to keep the thermal budget as low as possible, the 900°C-30 min annealing condition is a practical compromise, as only 50 mV of additional voltage is added for the 105 × 105 μm$^2$ device at 100 A/cm$^2$. For the 900°C 30 min sample, the J-V characteristics of different device sizes are plotted in Fig. 4(b), showing good convergence of the J-V curves. This indicates uniform Mg doping distribution across all the devices from the optimized activation process. A slight deviation of current density among the devices at a voltage of > 4 V could be attributed to the self-heating of the test devices.

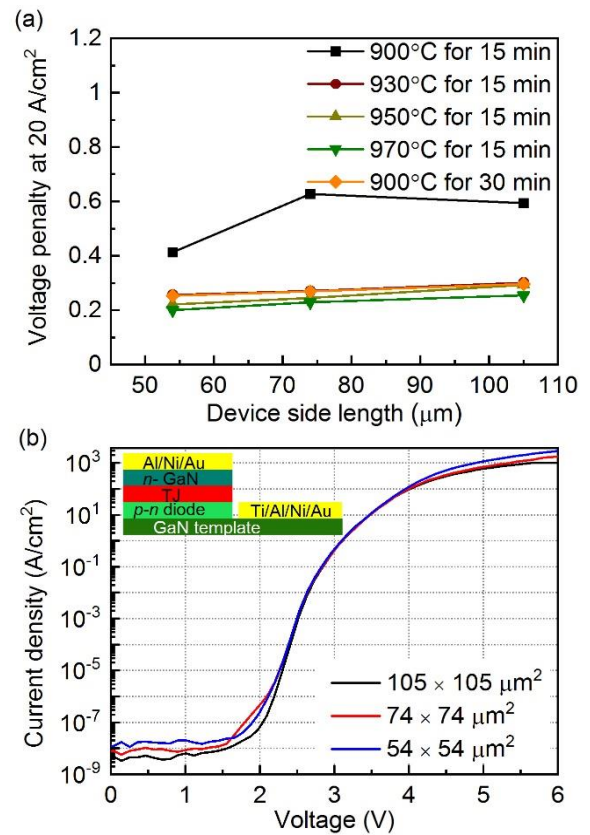

**Figure 4.** (a) Voltage drop at 20 A/cm$^2$ as a function of side length activated at different annealing conditions, and (b) current density-voltage characteristics activated at the optimized condition "900°C-30 min" for the square-shaped devices.

*3.2 Delta-dose*

The effects of δ-dose on TJs and the resulting electrical properties were investigated next. For that, our optimized activation process was employed to another set of samples with different δ-doses in the TJs. In addition to improving electrical conductivity, the δ-dose is expected to significantly impact an interface morphology. As a matter of fact, with an introduction of Mg and Si δ-dose on materials, a reduction in defect density by a factor of ×10 and ×2, respectively, was previously reported [36, 37]. This can be possibly understood through a growth interruption required during the δ-dose introduction, when dislocation density propagation can be partially terminated [37]. Moreover, the Mg δ-dose influences the reduction of the impurity self-compensation and enhancement of free carrier concentration for GaN and AlGaN.[37]

Considering the aforementioned benefits, Mg-δ samples with two surface densities of 2.2 × 10$^{13}$ cm$^{-2}$ and 4.5 × 10$^{13}$ cm$^{-2}$ (Fig. 1b) and a Si δ-dose sample with a surface density of 3 × 10$^{13}$ cm$^{-2}$ (Fig. 1c) were grown, fabricated and tested. To evaluate the diode performance, the voltage and normalized differential resistance against current density are plotted on a linear scale as shown in Fig. 5(a). With the incorporation of δ-dose in the devices, our device performance metrics, the voltage penalties are observed to decrease, and for a 4.5 × 10$^{13}$ cm$^{-2}$ Mg δ-dose sample, the voltage penalty is only 158 mV at 20 A/cm$^2$ and 490 mV at 100 A/cm$^2$ (Fig. 5b).

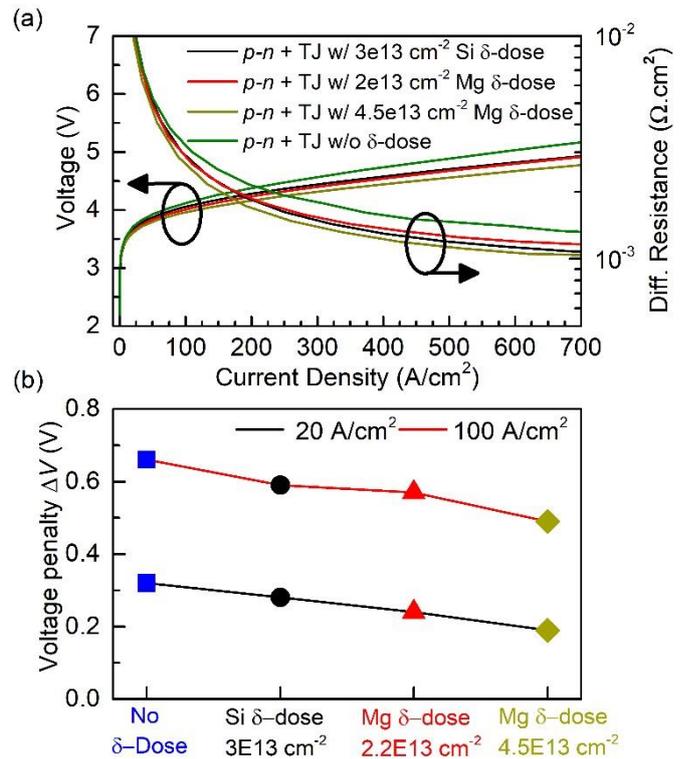

**Figure 5.** (a) J-V and normalized differential resistance-V characteristics, and (b) voltage penalty of samples no-δ, Si- δ, Mg- δ dose at 20 A/cm$^2$ and 100 A/cm$^2$.

The lowest normalized differential resistance among the TJ samples measured so far is 6.6 mΩ-cm2. This excellent result is achieved from the sample with a Mg-δ dose of 4.5 × 10$^{13}$ cm$^{-2}$, which is approximately half of the no-δ sample. Since all the p-n diodes underneath the TJs within the test devices have identical structures, the difference in the forward voltage drop is likely caused by the TJs. Both the theoretical and experimental studies show that implementing δ-dose at the interface improves the TJ efficiency by creating a thin, highly-doped layer which reduces the tunneling distance [38, 39]. The voltage penalty is reduced to 490 mV from 660 mV at 100 A/cm$^2$ as the Mg δ-dose with a concentration of $4.5 \times 10^{13}$ cm$^{-2}$ was introduced into the TJ. It should also be noted that the influence of Mg δ-dose on the TJ efficiency is higher compared to the Si δ-dose, as can be evidenced by the measured voltage penalties. Compare to the Si δ-dose sample, the Mg delta shows less penalty even for the relatively low doping concentration. The measured differential resistance is 3 mΩ-cm$^2$ at 100 A/cm$^2$. As the tunneling efficiency improves at high current density for a TJ, an increase of the input current density reduces the TJ resistance, and consequently, the overall resistance.

### 3.3 Mg buildup delay

While growing p-GaN structures with MOCVD, the Mg buildup delay [16] from the time Mg shutter is turned on, until the time required to reach the desired peak concentration has significant importance, especially while growing very thin layers required for TJs. To ensure maximum Mg incorporation in the highly p-doped layer, a series of samples of different p+-GaN layer thicknesses was tested. Interestingly, the diodes with a higher p+-GaN thickness exhibits a lower voltage drop, as shown in Fig. 6(a). Measuring the doping concentration or inspecting the surface morphology of the buried p-layers is challenging since all the samples under study were grown in a single-continuous growth. For the constant Mg δ-dose $4.5 \times 10^{13}$ cm$^{-2}$ and doping concentration of the p+ layer of the TJs, it is believed that with an increase of the p+ layer thickness, p-type conductivity improves and becomes maximum at a thickness of 12 nm. This suggests that 12-nm-thick p+-GaN layer results in highest tunneling efficiency due to the maximum achievable hole concentration. As can be seen in Fig. 6(b), further increase of the p-GaN thickness does not improve the electrical properties as the inherent tunneling distance supersedes the advantages of thicker heavily Mg-doped p+-GaN layers. Hence, growing the p+-GaN layer up to a certain thickness helps compensate the Mg buildup delay effect, resulting a low resistive TJ.

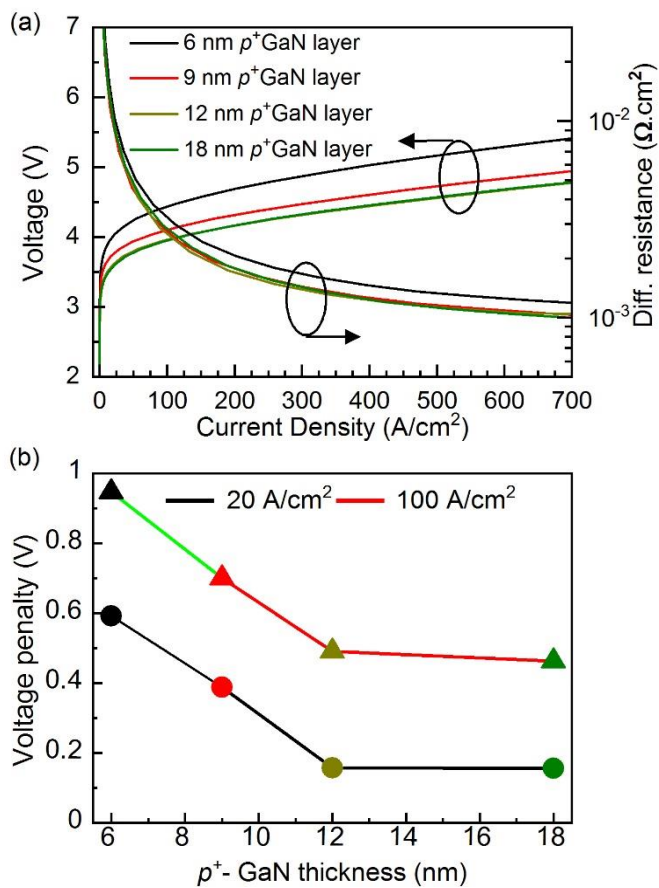

**Figure 6.** (a) J- and differential resistance against V for samples with different p+ layer thickness at the TJ, and (b) voltage penalty as a function of thickness of p+-GaN at 20 A/cm$^2$ and 100 A/cm$^2$.

To evaluate the TJ performance at a high current density than reported earlier for applications beyond optoelectronic and low-power electronic devices, one additional sample was processed for the Mg δ-dose sample (Fig. 1(b)). Figure 7 shows the *J*- and normalized differential resistance against *V* characteristics. The devices shows a record-low differential resistance of $1.6 \times 10^{-4}$ Ω-cm$^2$ at 5 kA/cm$^2$ which is the lowest value ever reported [23] for all-MOCVD grown GaN TJ devices. After the device turns on, the differential resistance drops until 5 kA/cm$^2$ and stays nearly constant in the current range from 5 kA/cm$^2$ – 10 kA/cm$^2$.

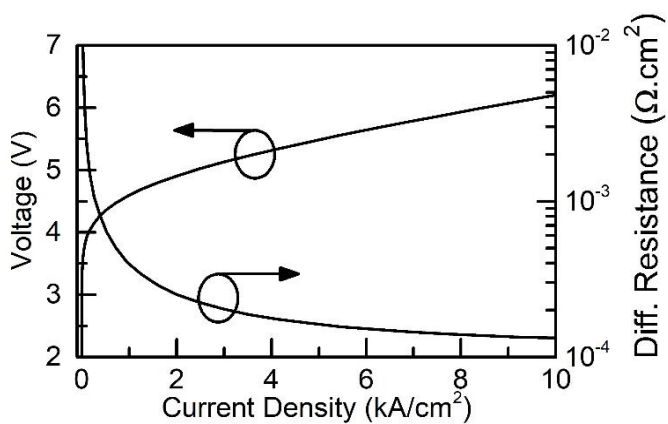

**Figure 7.** Current density and differential resistance vs voltage characteristics for the Mg-δ dose ($4.5 \times 10^{13}$ cm$^{-2}$) sample up to the current density of 10 kA/cm$^2$.

### 4. Conclusion

In conclusion, low resistivity TJs and reduced voltage penalties are obtained in all-MOCVD grown p-n diodes with a buried TJ. This was achieved by optimizing the activation annealing time and temperature for buried p-GaN layers. Additional improvement is also observed by implementing δ-dose for both donor and acceptor dopants within the TJs. Annealing at this high temperature is explored for the first time to our knowledge to achieve a uniform p-GaN condition that possibly opens a path for large-area all-MOCVD grown TJ integrated devices in the future. It is expected that combining both the Si and Mg δ-dose in an actual LED structure and activating the Mg-doped layer will result in further improvement in electrical performance, benefitting next-generation of electronic and optoelectronic devices.

### Acknowledgements

This material is based upon work supported by the U.S. Department of Energy's Office of Energy Efficiency and Renewable Energy (EERE) under the Building Technologies Office (BTO) Award Number 31150. Sandia National Laboratories is a multi-mission laboratory managed and operated by National Technology and Engineering Solutions of Sandia, LLC., a wholly owned subsidiary of Honeywell International, Inc., for the U.S. Department of Energy's National Nuclear Security Administration under contract DE-NA-0003525. The views expressed in the article do not necessarily represent the views of the U.S. Department of Energy or the United States Government.